\documentclass{elsart3p}
\usepackage{graphicx}
\usepackage{amssymb}

\begin{document}

\begin{frontmatter}

\title{Dynamics of kicked matter-wave solitons in an optical lattice}

\author{A. Cetoli$^{1}$, L. Salasnich$^{2}$,
B.A. Malomed$^{3}$, F. Toigo$^{4}$}

\address{$^1$Department of Physics,
Ume\aa ~University, SE-90187 Ume\aa , Sweden \\
$^{2}$CNISM and CNR-INFM, Unit\`a di Padova, Dipartimento di Fisica
``Galileo Galilei'', Universit\`a di Padova, Via Marzolo 8, 35131 Padova,
Italy \\
$^{3}$Department of Interdisciplinary Studies, School of Electrical
Engineering, Faculty of Engineering, Tel Aviv University, Tel Aviv 69978,
Israel \\
$^{4}$Dipartimento di Fisica ``Galileo Galilei'' and CNISM, Universit\`a di
Padova, Via Marzolo 8, 35131 Padova, Italy}

\begin{abstract}
We investigate effects of the application of a
kick to one-dimensional matter-wave solitons in a self-attractive
Bose-Einstein condensate trapped in a optical lattice. The
resulting soliton's dynamics is studied within the framework of the
time-dependent nonpolynomial Schr\"{o}dinger equation. The crossover
from the pinning to quasi-free motion crucially depends on the size
of the kick, strength of the self-attraction, and
parameters of the optical lattice.
\end{abstract}


\end{frontmatter}


\section{Introduction}

It is well known that Bose-Einstein condensates (BECs) with repulsive
inter-atomic interactions, characterized by a positive scattering length, $%
a_{s}>0$, can give rise to stable localized matter-wave states in the form
of \textit{gap solitons}, in the presence of an optical lattice (OL), which
induces a periodic potential acting on atoms. The existence of gap solitons
in BEC was predicted theoretically \cite{konotop} and demonstrated
experimentally \cite{ober}, see also review \cite{Markus}. Gap solitons are
represented by stationary solutions to the respective Gross-Pitaevskii
equation (GPE), with the eigenvalue (chemical potential) located in a finite
bandgap of the OL-induced spectrum. Note that a periodic potential emulating
the OL can also be induced by the spatially-periodic modulation of the
transverse trap which confines the condensate in the transverse directions
\cite{Napoli,sala-periodic2}.

In the limit of a very deep OL, the underlying GPE can be reduced to an
effective 1D discrete nonlinear Schr\"{o}dinger equation with the repulsive
on-site nonlinearity, assuming that the condensate's wave function is a
superposition of functions localized at particular lattice sites, with a
nearly vanishing overlap between them \cite{trombettoni}. In this limit, gap
solitons go over into \textit{staggered} discrete solitons, which feature
the alternation of the sign of the wave field between adjacent sites of the
lattice \cite{Panos}. On the other hand, if the corresponding OL is weak,
the general wave function may be split into a superposition of right- and
left-traveling modes, giving rise to a system of coupled-mode equations for
the respective slowly varying amplitudes (see, e.g., Ref. \cite{sakaguchi}),
which is tantamount to the well-known coupled-mode equations for optical
waves propagating through the Bragg grating \cite{optics}. Actually, gap
solitons were first predicted \cite{Aceves} and experimentally created \cite%
{Eggleton} as optical pulses in fiber Bragg gratings.

While the gap solitons are typically studied as standing localized states, a
detailed numerical analysis has demonstrated that an initial kick
(represented by a linear phase profile imprinted into the soliton) may set
that them in persistent motion, if the norm of the soliton (the number of
atoms bound in it) is below a certain threshold value. The existence of
moving gap solitons was predicted in both one- and two-dimensional (1D and
2D) settings \cite{sakaguchi}.

In the BEC with attractive interactions ($a_{s}<0$), solitons realize the
ground state of the condensate. Such solitons were created in condensates of
$^{7}$Li \cite{rice} and $^{85}$Rb \cite{Cornish} atoms, with the sign of
the atomic interactions switched to attraction by means of the
Feshbach-resonance technique (in the latter case, the solitons were observed
in a post-collapse state of the condensate). In the presence of a periodic
potential, such solitons should exist too, with the chemical potential
falling in the semi-infinite gap of the spectrum, as first shown in the
context of the optical setting \cite{Wang}, and later demonstrated in detail
in the framework of GPEs \cite{Alfimov,sala-periodic,sala-periodic2}.

The objective of this work is to predict possible regimes of motion of the
1D matter-wave solitons in the \emph{self-attractive} condensate, in the
presence of the OL potential. This will be done in the framework of the
nonpolynomial Schr\"{o}dinger equation (NPSE), which is known to provide for
high accuracy in the description of dynamical properties of solitons
supported by the self-attraction. This model is described in Section II,
while Section III summarizes the findings. The solitons will be first found
as stationary solutions to the NPSE, and then set in motion by suddenly
kicking them. Three dynamical regimes will be identified by systematic
simulations: steady motion, gradual decay of the moving solitons, and firm
pinning, when the kick cannot essentially disturb the initial soliton. In
particular, regions where these regimes occur will be charted in a parameter
plane of the model. The paper is concluded by a brief summary in Section IV.

\section{Bright solitons in the optical lattice}

Dynamics of atomic matter waves in rarefied BEC is very accurately described
by the time-dependent three-dimensional GPE,
\begin{equation}
i\hbar {\frac{\partial }{\partial t}}\psi =\left[ -{\frac{\hbar ^{2}}{2m}}%
\nabla ^{2}+U(\mathbf{r})+{\frac{4\pi \hbar ^{2}a_{s}}{m}}|\psi |^{2}\right]
\psi \;,  \label{GPE}
\end{equation}%
where $\psi (\mathbf{r},t)$ is the macroscopic wave function of the
condensate (normalized to the total number of atoms, $N$), $m$ the atomic
mass, and $a_{s}$ the $s$-wave scattering length of the inter-atomic
potential. As the trapping potential, we adopt the usual combination of the
tight confinement in the radial direction and longitudinal OL,
\begin{equation}
U(\mathbf{r})={(1/2)}m\omega _{\bot }^{2}(x^{2}+y^{2})+V(z),
\label{potential}
\end{equation}%
\begin{equation}
V(z)=-V_{0}\cos {(2k_{L}z)},  \label{potential-z}
\end{equation}%
with $k_{L}=2\pi /\lambda $, where $\lambda $ is the wavelength of
counterpropagating laser beams whose interference creates the OL, and $V_{0}$
is its effective depth. In the subsequent analysis, we use the effective
one-dimensional NPSE, which can be derived from Eq. (\ref{GPE}) by averaging
in the transverse plane \cite{sala-npse}. The NPSE has been demonstrated to
be very accurate in reproducing results that can be obtained from the full
3D GPE; in particular, this approach has been tested for bright \cite%
{sala-npse-bright} and dark solitons \cite{sala-npse-dark}, two-component
condensates \cite{sala-npse-2components}, and for the condensate in a
toroidal trap \cite{sala-npse-ring}, as well as for states with axial
vorticity \cite{sala-npse-vortex}. A similar approach, which generalizes the
Thomas-Fermi approximation and may give still more accurate results, but
does not apply to solitons, was later developed for tightly confined
self-repulsive condensates \cite{Delgado}.

The NPSE including axial OL potential (\ref{potential-z}) was derived too
\cite{sala-periodic,sala-periodic2}, assuming the following factorization of
the 3D wave function in Eq. (\ref{GPE}),%
\begin{equation}
\psi (\mathbf{r})=\exp {\left\{ -{\frac{\left( x^{2}+y^{2}\right) }{2\sqrt{%
1-g|f(z)|^{2}}}}\right\} }\frac{\,f(z)}{\left( 1-g|f(z)|^{2}\right) ^{1/4}},
\label{S}
\end{equation}%
where 1D wave function $f(z,t)$ is subject to normalization
\begin{equation}
\int |f(z)|^{2}\;dz=1,  \label{N=1}
\end{equation}%
the adimensional interaction strength is $g=2\,|a_{s}|N/a_{\perp }$, with $%
a_{\perp }=\sqrt{\hbar /(m\omega _{\perp })}$ the transverse trapping size.
The respective NPSE takes the form
\begin{eqnarray}
i\,\frac{\partial f(z,t)}{\partial t} &=&\left[ -{\frac{1}{2}}{\frac{%
\partial ^{2}}{\partial z^{2}}}+V(z)\right.   \nonumber \\
&&\left. +{\frac{1-{(3/2)}g|f(z,t)|^{2}}{\sqrt{1-g|f(z,t)|^{2}}}}\right]
f(z,t),  \label{f}
\end{eqnarray}%
with length and energy measured in units of $a_{\perp }$ and $\hbar \omega
_{\perp }$, respectively.

If applied to the ring-like (toroidal) geometry, Eq. (\ref{f}) predicts that
there exists a critical value, $g_{\min }$, above which the axially uniform
state in the torus becomes modulationally unstable, and is replaced, as the
ground state, by a soliton-like configuration. Moreover, there is another
critical value of the interaction strength, $g_{\mathrm{coll}}$, above which
the localized solution ceases to exist due to the collapse, hence the
soliton persists in interval $g_{\min }<g<g_{\mathrm{coll}}$. Similar
effects in the toroidal setting were earlier studied in Refs. \cite{carr}.

Both the full three-dimensional GPE and its NPSE reduction are mean-field
equations. Contrary to the repulsive case, in the attractive case that we
are dealing with, there is no superfluid-Mott insulator transition.
Beyond-mean-field effects are irrelevant in our case too, since, as shown in
Ref. \cite{links}, they appear at lattice sites populated with a small
number of particles ($\leq 5$). In our case, due to the attractive
interaction, most of the atoms reside in few highly populated sites.

As concerns quantum effects, it may be relevant to mention that, unlike
regular solitons in attractive condensates, gap solitons in the BEC with
repulsion cannot represent the ground state. Therefore, in the framework of
the full quantum description, the stability of such states needs further
analysis. In particular, it was demonstrated that dark solitons in BEC with
repulsive interactions between atoms are subject to quantum depletion, with
incoherent atoms gradually filling the dark-soliton's notch under the action
of anomalous fluctuations \cite{Dziarmaga}. Nevertheless, dark solitons,
which accurately obey the mean-field description provided by the GPE, were
successfully observed in the experiment \cite{dark-soliton}. Because loosely
bound gap solitons may feature a set of notches in their \textquotedblleft
tails", they may also be subject to the quantum depletion, in that sense. In
terms of the quantum theory, this remains an open question, but the
experimental results \cite{ober,Markus} clearly demonstrate that gap
solitons can be created, following the predictions of the mean-field theory.
On the other hand, for tightly bound soliton in the attractive BEC (as well
as for tightly gap solitons in the repulsive condensate, if they are created
not too close to edges of the respective bandgap \cite{SKA}), the quantum
depletion should not be an issue, as their existence is not predicated on
the presence of notches.

\section{Phase imprinting and ensuing dynamics}

As said above, the objective of this work is to analyze the dynamics of
moving bright solitons in the model introduced above, which includes the
periodic OL potential. The motion will be initiated by sudden application to
the soliton of a kick with momentum $p$.

First, we generate a stationary soliton by numerically solving the
corresponding stationary equation, obtained from Eq. (\ref{f}) by the
substitution of $f(z,t)=e^{-i\mu t}\phi (z)$, where $\mu $ is the real
chemical potential and $\phi (z)$ a real stationary wave function:
\begin{equation}
\left[ -{\frac{1}{2}}{\frac{d^{2}}{dz^{2}}}+V(z)+{\frac{1-{(3/2)}g\phi ^{2}}{%
\sqrt{1-g\phi ^{2}}}}\right] \phi (z)=\mu \,\phi (z).  \label{static-npse}
\end{equation}

We fix $k_{L}\equiv 1$ in potential (\ref{potential-z}), and solve Eq. (\ref%
{static-npse}) with periodic boundary conditions, $f(z)\equiv f(z+L)$, with
the total length of the ring-shaped system equivalent to $32$ periods of the
OL, $L=32\pi $. Taking $V_{0}=0.5$ for the OL depth, we find the ground
state of Eq. (\ref{static-npse}) by means of the numerical scheme described
in Refs. \cite{sala-periodic,sala-periodic2}, for values of the interaction
strength $g$ belonging to the above-mentioned existence range, $g_{\min
}<g<g_{\mathrm{coll}}$.

To initiate the dynamics, we multiply the so determined stationary solution $%
\phi (z)$ by $\exp (ipz)$, i.e., use the following initial configuration,
\begin{equation}
f(z,t=0)=\phi (z)\exp (ipz)\;,  \label{wave-kick}
\end{equation}%
where momentum $p$ is compatible with the periodic boundary conditions,
taking values $p=n/16$ with integer $n$. We pick up such values in interval $%
0<p<0.5$. Configuration (\ref{wave-kick}) can be created in the experiment
by means of the well-known phase-imprinting technique \cite{phase-imprinting}%
. The full time-dependent NPSE, Eq. (\ref{f}), with initial condition (\ref%
{wave-kick}), was solved using the Crank-Nicholson predictor-corrector
algorithm in real time \cite{sala-numerics}.

\begin{figure}[tbp]
\centering
\includegraphics[height=6.15in,clip]{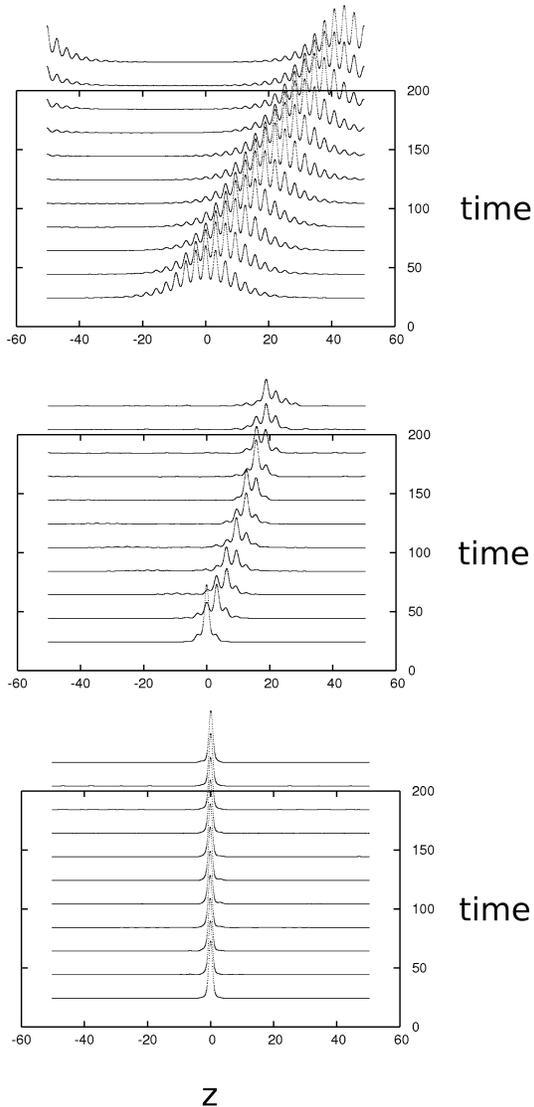}
\caption{Profiles $|f(z,t)|^{2}$ observed in the course of the evolution of
the kicked soliton, with initial momentum $p=0.25$. From top to bottom: $%
g=0.15$, $g=0.5$ and $g=0.8$. Here and in other figures, parameters of the
optical lattice are $k_{L}=1$ and (unless specified otherwise) $V_{0}=0.5$.}
\label{solirun-f1}
\end{figure}

Examples of the simulated evolution are displayed in Fig. \ref{solirun-f1},
which suggests the existence of different regimes corresponding to different
values of interaction strength $g$. In particular, for small $g$ ($g=0.15$),
the stationary soliton extends over many lattice cells and, after receiving
the kick, it moves in quite a regular way. At intermediate values of the
interaction strength, $g=0.5$, the soliton occupies a few sites, and its
dynamics generated by the action of the kick is irregular. For a larger
strength, $g=0.8$, the initial wave packet is essentially localized in a
single cell of the lattice, and stays trapped in the same cell (for the
entire period of our simulation) after the application of the kick.

\begin{figure}[tbp]
\centering
\includegraphics[height=2.3in,clip]{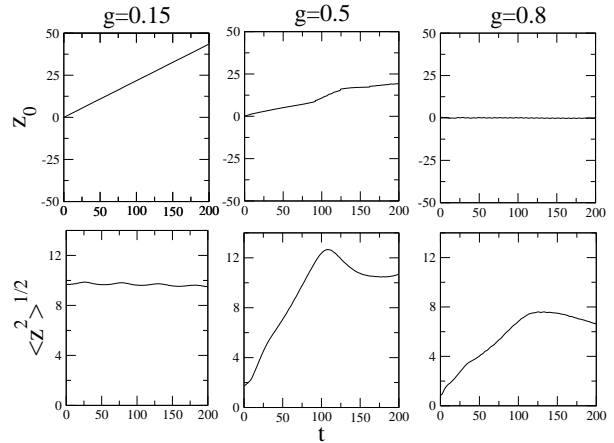}
\caption{The center of mass, $z_{0}$, and average axial width, $\langle
z^{2}\rangle ^{1/2}$, of the soliton as functions of time. The initial
momentum is $p=0.25$.}
\label{solirun-f2}
\end{figure}

To characterize the motion of the kicked soliton in a more accurate form, we
calculate the average axial position of the soliton, $z_{0}(t)=\int dz\ z\
\left\vert f(z,t)\right\vert ^{2}$, and its average squared width,
\begin{equation}
\langle z^{2}(t)\rangle =\int dz\ (z-z_{0}(t))^{2}\ \left\vert
f(z,t)\right\vert ^{2},  \label{mean}
\end{equation}%
where the integration is performed over spatial period $L\equiv 32\pi $ (the
expressions are not divided by the norm of the wave function because it is
fixed to be $1$, as per Eq. (\ref{N=1})). These characteristics are shown,
as functions of time, in Fig. \ref{solirun-f2} for the same values of $g$ as
in Fig. \ref{solirun-f1}. The figure shows that, with $g=0.15$, the
soliton's center of mass ($z_{0}$) moves at a constant speed. We have
verified that the respective speed is well fitted by expression $v=p/m^{\ast
}$, with the effective inverse mass of the ground state, $1/m^{\ast
}=1-V_{0}^{2}/2$, as predicted by the usual theory of the quasi-momentum in
periodic lattices \cite{ashcroft}. Simultaneously, the width of the soliton,
$\sqrt{\langle z^{2}\rangle }$, displays periodic small-amplitude
oscillations, i.e., the moving soliton behaves as a \textit{breather}. At $%
g=0.5$, the motion of the soliton is less regular: its velocity is not
constant, and the width increases in the course of the evolution, i.e., the
soliton gradually spreads out; however, after becoming much broader than it
was initially (by a factor $\sim 6$), it seems to stabilize itself. At $g=0.8
$, the initial width of the soliton, $\sqrt{\left\langle
z^{2}(t=0)\right\rangle }$, is, as said above, comparable to the size of the
lattice cell. In this case, the soliton's center of mass does not exhibit
any systematic motion, while the width initially increases (by a factor $%
\sim 4$), but then ceases to grow.

\begin{figure}[tbp]
\centering\includegraphics[height=1.8in,clip]{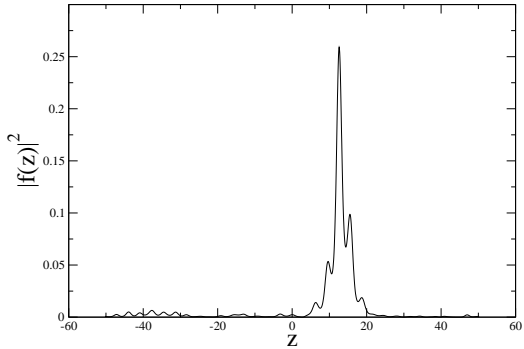}
\caption{Snapshot of $|f(z,t)|^{2}$ at $t=108$ for the soliton in the
central panel of Fig. \protect\ref{solirun-f1}, which corresponds to the
largest value of $\langle z^{2}\rangle ^{1/2}$}
\label{solirun-f3}
\end{figure}

Thus, we do not observe complete delocalization in the regime of spreading .
Instead, it is concluded that the soliton interacts with the OL in a complex
way, emitting small-amplitude waves, which affect the time evolution of the
average width. Figure \ref{solirun-f3} shows the soliton's profile at $t=108$
in the case corresponding to the central panel of Fig. \ref{solirun-f1},
which features the maximum value of $\sqrt{\langle z^{2}\rangle }$. The
figure demonstrates the presence of \textquotedblleft splinters", to the
left of the soliton, which actually break off from the soliton shortly after
the application of the initial kick. They travel backwards with respect to
the soliton, reaching the largest distance from it at $t=108$, due to the
periodic boundary conditions. In this situation, expression (\ref{mean})
does not provide an adequate measure to estimate the spreading of the
soliton. In principle, after performing a round trip in the toroidal trap,
the splinters may hit the soliton, but the study of this issue require
extremely long simulations, which we did not perform in the framework of the
present work.

As a better measure of the spreading, we introduce the following \textit{%
effective entropy},
\begin{equation}
S(t)=-\sum_{n}A_{n}(t)\frac{\ln {A_{n}(t)}}{\ln (N_{\mathrm{cell}})}~,
\label{entropy}
\end{equation}%
where $A_{n}(t)$ is the share of the norm located, at time $t$, within the $n
$-th lattice cell, i.e. $A_{n}(t)=\int_{\pi n}^{\pi (n+1)}dz|f(z,t)|^{2}$,
and $N_{\mathrm{cell}}$ is the total number of cells [recall we run the
simulations for $N_{\mathrm{cell}}=32$, and $\sum_{n=1}^{N_{\mathrm{cell}%
}}A_{n}=1$, due to normalization condition (\ref{N=1})]. Definition (\ref%
{entropy}) yields the maximum of the entropy, $S=1$, if the matter is
distributed uniformly, $f(z,t)=\mathrm{const}$, and the entropy attains its
minimum, $S=0$, if the entire norm is concentrated in a single cell.

We have found that the unnormalized effective entropy, i.e. $S(t)\ln (N_{%
\mathrm{cell}})$, turns out to be practically independent of the total
length of the system in the axial direction. In particular, while the
mean-square width is affected by the backscattered radiation emitted after
the kick, the entropy is not. More precisely, we have checked that the
result of the computation of the unnormalized entropy does not change with
the variation of the total lengths.

\begin{figure}[tbp]
\centering\includegraphics[height=1.25in,clip]{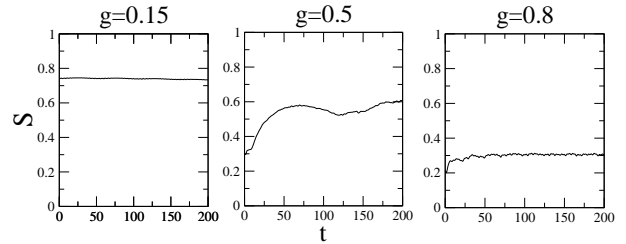}
\caption{The same cases as in Fig. \protect\ref{solirun-f2} are shown here
in terms of the evolution of entropy (\protect\ref{entropy}).}
\label{solirun-f4}
\end{figure}

In Fig. \ref{solirun-f4} we plot $S(t)$ as a function of time for the same
runs which were included in Fig. \ref{solirun-f2}. It is seen that, for $%
g=0.15$, initial entropy $S(0)$ is large, because the soliton covers many
cells, and, in the course of the evolution, $S(t)$ exhibits periodic
small-amplitude oscillations around $S(0)$. On the contrary, for $g=0.5$ the
stationary soliton covers few cells, hence initial entropy $S(0)$ is quite
small, but the entropy increases as a consequence of the soliton's spreading
out. For $g=0.8$, the initial entropy is similar to that in the case of $%
g=0.5$, but its evolution is completely different, featuring very small
aperiodic oscillations around $S(0)$, i.e. $S(t)$ remains nearly constant,
as the narrow strongly pinned soliton does not start conspicuous motion or
spreading out after being kicked.

\begin{figure}[tbp]
\centering
\includegraphics[height=2.in,clip]{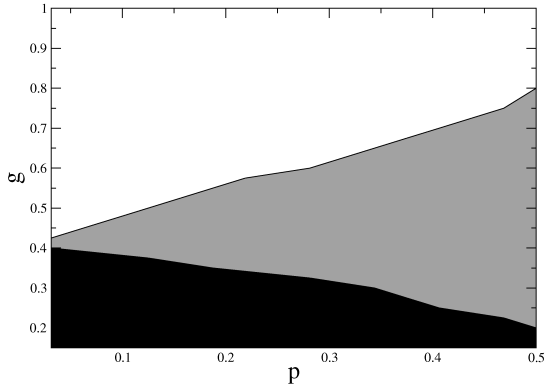}
\caption{The map of different dynamical regimes in the plane of $(p,g)$,
where $p$ is the momentum imparted to the soliton by the initial kick, and $g
$ the strength of the self-attractive nonlinearity. The black region:
steadily moving breather-like solitons; the gray region: spreading out of
irregularly moving solitons; the white region: firm pinning (the soliton
does not start progressive motion). 
Lattice parameters: $V_0=0.5$ and $k_L=1$.}
\label{solirun-f5}
\end{figure}

Thus, our systematic simulations reveal the existence of three different
dynamical regimes: (i) stable breathers, i.e., solitons steadily moving
across the lattice potential at an almost constant velocity, with
small-amplitude shape oscillations; (ii) dispersive dynamics, in which case
the soliton strongly spreads out in the course of the evolution; (iii)
localization, in which a narrow soliton remains trapped in one lattice cell.
In Fig. \ref{solirun-f5}, we plot the respective diagram in the parameter
plane of $(p,g)$, where these three regimes are mapped as follows.

(i) The black region represents steadily traveling solitons, whose effective
entropy varies by $\leq 5\%$ in the course of the long evolution.

(ii) The gray region: the dispersive regime, in which the solitons move at a
variable speed, and their effective entropy increases by more than $5\%$
against the initial value.

(iii) The white region: localized solitons, with the center of mass firmly
pinned to the initial position. In the latter case, the solitons may be
slightly dispersive at the initial stage of the evolution, but their
effective entropy is always much smaller than in the other two cases.

Figure \ref{solirun-f5} summarizes results of $560$ numerical runs, with $16$
values of $p$ and $35$ values of $g$. The momentum resolution is imposed by
the periodic boundary conditions: $\Delta p=2\pi /L$, where $L\equiv 32\pi $
is the total axial length, as defined above. By choosing larger $L$, one can
reduce $\Delta p$. The step in the variation of strength $g$ in Fig. \ref%
{solirun-f5} is $\Delta g=0.25$.

Figure \ref{solirun-f5} shows that the regime of the stable motion is
confined to the region of small values of interaction strength $g$. For 
fixed $g$, a transition from the steady motion to the dispersive regime is
observed with the increase of kick $p$, and the numerical results suggest 
that the steady-motion region disappears as $p$ attains large values. As
expected, the increase of $g$ drives the system into the regime of pinning,
but one can leave it by increasing the kick.

It should be stressed that, while the border between the localization and
other two regimes is well defined, the exact location of the boundary
between the steady-motion and dispersive regimes depends on our choice of
the $5\%$ maximum for the allowed change of the entropy in the course of the
evolution. Admitting a $10\%$ change does not conspicuously affects the
location of the border, but defining the threshold of $20\%$ makes the
region of the stable motion slightly larger (not shown here in detail).

At the boundary between motion and dispersion regimes, we observe a density
profile that becomes more and more noisy, while the velocity of the
traveling soliton decreases in an irregular way, similar to some regimes of
the motion of gap solitons in the 1D model with the repulsive cubic
nonlinearity, which were reported in Ref. \cite{sakaguchi}. This drop in the
velocity is also relevant to the identification of the boundary of the
pinning regime, where, typically, the kicked soliton visits a few lattice
sites and then stops. For instance, at $g=0.55$ and $k=0.25$, the soliton
passes a few sites, then bounces back to the original position, and stays
pinned there.

\begin{figure}[tbp]
\centering
\includegraphics[height=2.in,clip]{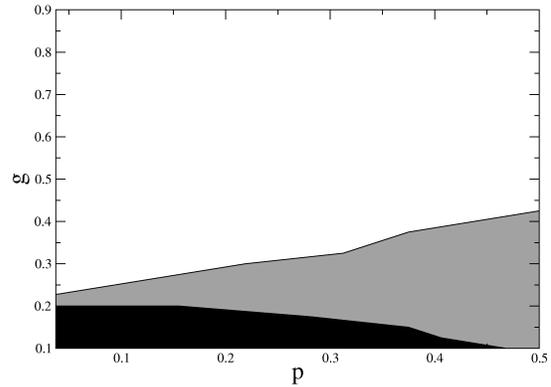}
\caption{The same as in Fig. \protect\ref{solirun-f5}, but for $V_{0}=1$.}
\label{solirun-f6}
\end{figure}

The diagram of the dynamical regimes depends on parameters of the optical
lattice, $V_{0}$ and $k_{L}$, see Eq. (\ref{potential-z}). In particular,
Fig. \ref{solirun-f6}, obtained by collecting results of other $16\times
33=528$ numerical runs, displays this diagram for a stronger lattice, with $%
V_{0}=1$, while the OL period remains equal to $\pi $ (i.e., $k_{L}=1$).
Qualitatively, the picture is similar to that shown in Fig. \ref{solirun-f5}%
, but there are apparent quantitative differences. In particular, the
regions of the stable motion and dispersive behavior are strongly reduced,
which is not surprising, as the deeper OL can pin the solitons stronger. In
addition, the figure shows that there exists a critical momentum, $p_{c}$,
above which the region of steady motion does not exist anymore. This
critical value $p_{c}$ decreases with the increase of OL depth $V_{0}$.

As mentioned above, the upper bound for the considered values of the
interaction strength, $g$, is imposed by the onset of the collapse, $g=g_{%
\mathrm{coll}}$. In particular, in Ref. \cite{sala-periodic} it was found
that $g_{c}=1.07$ for $V_{0}=0.5$, and $g_{\mathrm{coll}}=0.96$ for $V_{0}=1$
(if $k_{L}=1$ is fixed). On the other hand, the minimum strength above which
soliton states exist is $g_{\min }\sim \pi ^{2}/L$ \cite{carr}; in this
work, this border does not manifest itself, as we considered values of $g$
at which the soliton is well localized within the lattice; in fact, we
always took $\sqrt{\langle z^{2}(0)\rangle }<\,L/10$.

In an attempt to understand the behavior of the kicked soliton in analytical
terms, we tried a variational approach. Using a Gaussian \textit{ansatz}, we
derived the respective variational equations of motion. We have thus
obtained a set of equations equivalent to that recently reported in Ref.
\cite{berry-kutz}. However, a careful analysis demonstrates that this
approach fails to describe the soliton's dynamics correctly in the present
model: although it does predict regions of pinning, moving breathers, and
solutions whose width expands indefinitely, the resulting diagram is quite
different from those in Figs. \ref{solirun-f5} and \ref{solirun-f6}. This
problem might be expected, since, even in the static case, the Gaussian
ansatz does not reproduce the behavior revealed by numerical solutions \cite%
{sala-periodic}.

\section{Conclusions}

We have studied the dynamics of moving matter-wave solitons in the
self-attractive BEC trapped in the axial optical-lattice potential. The
dynamics was induced by kicking a stationary soliton via the phase
imprinting. We have performed systematic simulations of the time-dependent
nonpolynomial Schr\"{o}dinger equation, which accurately describes
Bose-Einstein condensates under the tight transverse confinement.

Three dynamical regimes have been identified. The first of them is the
steady motion of stable solitons with small-amplitude intrinsic
oscillations, which is observed at relatively small values of the strength
of the inter-atomic attraction $g$, and relatively small size of the kick $p$%
, and the second is the dispersive regime, in which the soliton's velocity
decreases irregularly, while the soliton suffers systematic spreading. The
latter outcome of the application of the kick to the soliton also occurs for
relatively small $g$, but at larger $p$. The third regime naturally features
firm pinning, at large values of $g$ and small $p$. A somewhat similar
description for dynamical regimes was elaborated in Ref. \cite{trombettoni},
but the position of the different regions in the respective parameter plane
is totally different, due to the fact that the quasi-discrete limit
considered in that work is not adequate for comparison with the present
model.

Available experimental techniques \cite{rice,Cornish} make the experimental
realization of the predicted dynamical regimes quite feasible. It may also
be relevant, for the purposes of the theory and plausible experiment alike,
to study in details collisions between steadily moving solitons. The latter
problem will be considered elsewhere.

\section*{Acknowledgements}

This work has been partially supported by Fondazione CARIPARO. B.A.M.
appreciates hospitality of the Department of Physics ``Galileo Galilei'' at
the University of Padova (Padua). L.S. thanks GNFM-INdAM for partial support.

\end{document}